\documentclass[english,aps, two column]{revtex4}
\usepackage[latin1]{inputenc}
\usepackage{graphicx}

\makeatletter

 \newenvironment{lyxlist}[1]
   {\begin{list}{}
     {\settowidth{\labelwidth}{#1}
      \setlength{\leftmargin}{\labelwidth}
      \addtolength{\leftmargin}{\labelsep}
      }}
   {\end{list}}

\usepackage{babel}
\makeatother
\begin{document}

\preprint{Preprint}

\title{Microwave Irradiation Effects on Random Telegraph Signal in a MOSFET }

\author{Enrico Prati}

\affiliation{Laboratorio Nazionale Materiali e Dispositivi per la Microelettronica,
Consiglio Nazionale delle Ricerche - Istituto Nazionale per la Fisica della Materia, Via Olivetti 2, I-20041
Agrate Brianza, Italy}

\author{Marco Fanciulli}

\email{marco.fanciulli@mdm.infm.it}

\affiliation{Laboratorio Nazionale Materiali e Dispositivi per la Microelettronica,
Consiglio Nazionale delle Ricerche - Istituto Nazionale per la Fisica della Materia, Via Olivetti 2, I-20041
Agrate Brianza, Italy}

\author{Alessandro Calderoni}

\affiliation{Dipartimento di Elettronica e Informazione, Politecnico di Milano,
P.za Leonardo da Vinci 32, I-20133 Milano, Italy}

\author{Giorgio Ferrari}

\affiliation{Dipartimento di Elettronica e Informazione, Politecnico di Milano,
P.za Leonardo da Vinci 32, I-20133 Milano, Italy}

\author{Marco Sampietro}

\affiliation{Dipartimento di Elettronica e Informazione, Politecnico di Milano,
P.za Leonardo da Vinci 32, I-20133 Milano, Italy}

\begin{abstract}
We report on the change of the characteristic times of the random
telegraph signal (RTS) in a MOSFET operated
under microwave irradiation up to 40 GHz as the microwave field power is raised. The effect is explained by
considering the time dependency of the transition probabilities due
to a harmonic voltage generated by the microwave field that couples
with the wires connecting the MOSFET. From the dc current excited into the MOSFET by the
microwave field we determine the corresponding equivalent
drain voltage. The RTS experimental
data are in agreement with the prediction obtained with the model, making
use of the voltage data measured with the independent dc microwave
induced current. We conclude that when operating a MOSFET
under microwave irradiation, as in single spin resonance detection,
one has to pay attention into the effects related to microwave irradiation 
dependent RTS changes.
\end{abstract}

\maketitle
The Random Telegraph Signal (RTS), observed as a random
switching between two states of the channel current in a metal-oxide-semiconductor
field effect transistor (MOSFET) \cite{Ralls 86,Kandia 89,Simoen 02,Uren 85},
has been considered as a possible quantum readout mechanism by Vrijen
et al. \cite{Vrijen 00} and has been used for detecting single
electron spin resonance.\cite{Xiao 04} The spin resonance detection requires
the irradiation by a microwave field of a defect at the
$Si/SiO{}_{2}$ interface of a MOSFET in presence of a static magnetic field.\cite{Martin 03,Prati 06}
The capture $\lambda_{c}$ and emission $\lambda_{e}$ rates
due to the tunneling of electrons assisted by multiphonon non radiative processes
depend on the energy levels of the trap with respect to the Fermi energy
of the electron channel.\cite{Palma 97} The change in the rates, at the resonance frequency,
is due to the microwave-induced transition between the Zeeman energy levels of the 
trap. The RTS change at the spin resonance is detected by monitoring either
the average drain-source current or the emission and capture times
as a function of the static magnetic field while irradiating the device with a fixed microwave field.\cite{Xiao 04}

The RTS change in spin resonance condition is detected by monitoring either
the average drain-source current, or the emission and capture times
as a function of the static magnetic field while irradiating the device with a microwave field.
In both cases one should carefully identify proper experimental conditions in order to avoid 
spurious resonances induced by other traps in the MOSFET. We have already shown that the average
current method is affected by a microwave induced stationary drain-source
current.\cite{Ferrari 05} 

In this letter we demonstrate that also the
emission and capture times of the trap may change as a function of
the intensity of the microwave field. To this aim, we have
systematically characterized a change of the mean emission time
$\tau_{e}$ and capture time $\tau_{c}$ of a trap at the interface between silicon and oxide in 
n-MOSFETs interacting with a microwave field. The devices are made
on a p-well, with channel length of 0.35 $\mu$m, width of 0.45 $\mu$m,
an oxide thickness of 7.6 nm, a threshold voltage of about 460
mV, and a transition frequency of several tens of GHz. All the contacts of source, drain, gate, and well were directly
accessible through the bonding pads and connected to wires. 
The current $I_{D}$ flowing through drain and source is measured by a
transimpedance amplifier whose output is sampled and digitized for
off-line processing. The bandwidth of the amplifier extends to about
240 kHz allowing to characterize traps down to few microsecond
characteristic times. The microwave source is a dipole antenna
placed in front of the device, operating in a broad frequency range
from 1 GHz to 40 GHz. The reported power refers to the power of the microwave generator at the source.

Figure 1 shows the variation of emission
(down triangles, low current state) and capture (up triangles, high current state) characteristic times of our sample,
in a given condition of MOSFET biasing, as a function of the
microwave power, the frequency remaining fixed at $\nu$=
15.26 GHz. The figure shows that in the trap under investigation 
the characteristic time $\tau_{c}$ is a function of the microwave power, while
$\tau_{e}$ is constant.

The effect shown in Figure 1 can be fully ascribed to the
inevitably present coupling between the microwave field and the
conductive loop formed by the MOSFET and the connections toward the
sensing amplifier: the microwave field induces a harmonic current on the loop, modulating
the source and drain voltages of the MOSFET. In circuital
representation, this corresponds to adding two AC voltage generators
at the drain and at the source of the MOSFET (see Figure 2) with the same frequency of the
microwave field. As shown in Figure 3, in static condition the characteristic times
$\tau_{e}$ and $\tau_{c}$ change as a function of $V_{D}$.
Exploiting the change of the field distribution in the proximity of
the dipole antenna as a function of the frequency of the microwave,
we are able to set a frequency where the coupling of the MOSFET with
the microwave field occurs only at the drain. To set such condition
we used a microwave frequency of 15.26 GHz. At room temperature, without
microwave field applied and at $V_{G}=800$ mV, the RTS has a mean capture
time $\tau_{c}$ ranging monotonically from 3 ms to 20 ms for a drain
voltage variation from 200 to 800 mV, while the mean emission time
$\tau_{e}$ remains almost constant at about 0.7 ms.

In order to correlate the results of Figure 3 with those of Figure 1 obtained 
with the drain-source voltage oscillating at the microwave frequency, we calculated from Figure 3 the 
instantaneous capture and emission
probabilities (per unit time), $\lambda_{c}$ and $\lambda_{e}$, as the inverse of
the mean times for any drain voltage:
$\lambda_{c}(V_{D})=1/\tau_c(V_{D})$ and
$\lambda_{e}(V_{D})=1/\tau_e(V_{D})$.\cite{Machlup 54} The
modulation of $V_{D}(t)$ induced by the microwave field implies a
modulation of the capture and emission probabilities 
$\lambda_{c,e}(V_{D}(t))$. Since the microwave frequency $\nu$ has a
period $T=1/\nu$ much shorter than the RTS characteristic times, we
assume that the capture and emission processes are controlled by the
average probabilities:
$\overline{\lambda_{c,e}}=1/T \int_{0}^{T}
\lambda_{c,e}(V_{D}(t)){\rm d}t$, where the integration is performed over the period T of the microwave field. Since 
generally $\lambda_{c,e}$ are not linear functions of $V_{DS}$, the average value calculated by integrating upon a period
differs from the value that the probability assumes when the amplitude of the harmonic voltage is zero.

In this framework, the characteristic times with a microwave field applied can be obtained
as $\tau_{c,e}=1/\overline{\lambda_{c,e}}$. To quantitatively determine the 
value of $\tau_{c,e}$ as a function of the microwave power, it is necessary to determine the amplitude $v_{ac}$ of 
the modulating voltage at the drain induced by the microwave field. 

To calculate such amplitude at each microwave power applied, we profit from the fact that a 
dc stationary current is also generated as the microwave field is raised due to the rectification produced by
the non-linear I-V characteristic of the MOSFET.\cite{Ferrari 05} The current can be
fitted by calculating the average of the current values as a function of the drain voltage around the bias
value of $V_{D}$, weighted properly on a period, and setting the voltage amplitude as the only free parameter to be determined. 
The voltage amplitude can now be converted into the nominal microwave power $P_{\mu
w}$. The experimental data of the dc current as a function of the microwave power are reported in Figure
4. From these data we obtain: $v_{ac}= 1.15 [V/W^{1/2}] \cdot \sqrt{P_{\mu w}}$. Such relationship is used to predict
the change of $\tau_{c}$ as a function of the microwave field power (Figure 1), in excellent agreement
 with the experiment. The constancy of $\tau_{e}$ (Figure 1) with respect to the microwave power
agrees with the independence of $\tau_{e}$ of $V_{D}$. Different traps may have an opposite behaviour if $\tau_{c}$ 
is independent of $V_{D}$ or a mixed one if both $\tau_{c}$ and $\tau_{e}$ depend on $V_{D}$.

Such an agreement between the RTS variations and the microwave power has been confirmed on a variety 
of experiments carried out on various samples held in significantly different experimental
conditions: samples inserted in resonant cavities operating at frequencies in X-band (9.5 GHz) and
Q-band (34 GHz), different temperatures from 1 K to 300 K and also hybrid conditions when both the
source and the drain are coupled to the microwave field.

To summarize, the inevitably present electric loop due to the on-chip and off-chip
connections of a MOSFET to the external measuring system is
responsible for the detected RTS variation upon
microwave irradiation through the modulation of the MOSFET biasing
conditions. Such an effect has important consequences on single-spin resonance experiments. A spurious microwave
absorption in the environment may vary the effective power of the
microwave field coupled with the MOSFET and produce a change of the RTS characteristics not related to 
the trap - responsible for the RTS - driven in spin resonance conditions.
Any measured $\tau_c$, $\tau_e$ and dc current change in agreement with the proposed model prediction at a given power 
absorption has to be regarded as a spurious effect and cannot be ascribed to a single spin resonance phenomenon.

\begin{acknowledgments}
The authors would like to thank Mario Alia (MDM-INFM) and Sergio
Masci (Politecnico di Milano) for the samples preparation.
\end{acknowledgments}

\section{Figure Captions}

\begin{lyxlist}{00.00.0000}
\item [Fig.1]Experimental mean capture (up triangles) and emission (down triangles) times versus microwave field power 
compared to the theoretical predictions (continuous lines). Experimental conditions: $V_G$= 800 mV, $V_D$= 500 mV, 
$\tau_{c0}$= 14 ms, $\tau_{e0}$= 0.7 ms, $\nu$=15.26 GHz.

\item [Fig.2]Electrical circuit of the loop formed by the MOSFET and the 
sensing amplifier (not irradiated by the microwave field) coupled to the microwave 
field. The voltage induced by the field is represented by the 
two voltage generators $v_{d\mu w}$ and $v_{s\mu w}$.

\item [Fig.3]Experimental capture (up triangle) and emission (down triangle) time constants as a function of the 
drain voltage for $V_G=800 mV$ and $V_S= 0V$.

\item [Fig.4]Measurement (circles) and simulation (continuous line) result of the dc drain current as a function 
of the microwave field power. Experimental conditions: $V_G$= 800 mV, $V_D$= 500 mV, $I_{DS0}$= 4.3 $\mu$A, $\nu$=15.26 GHz.
\end{lyxlist}

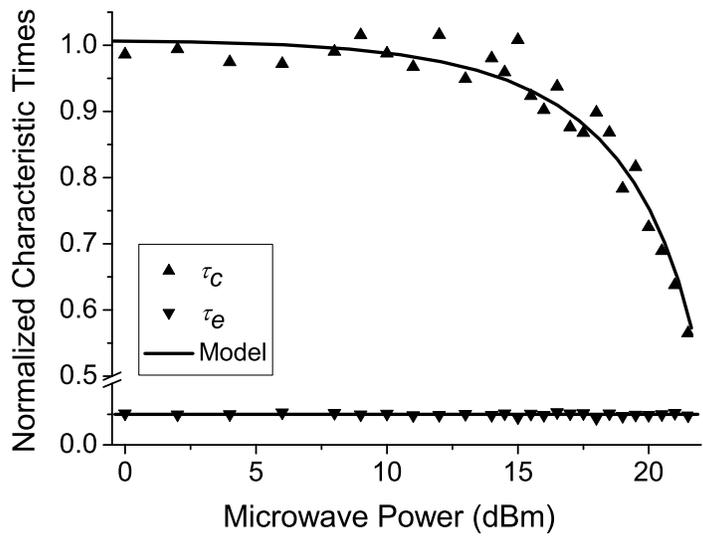
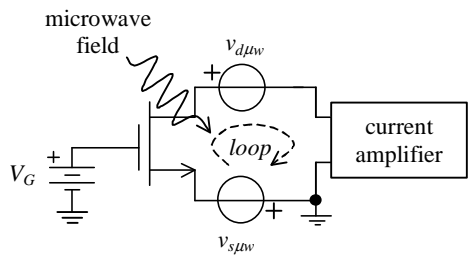
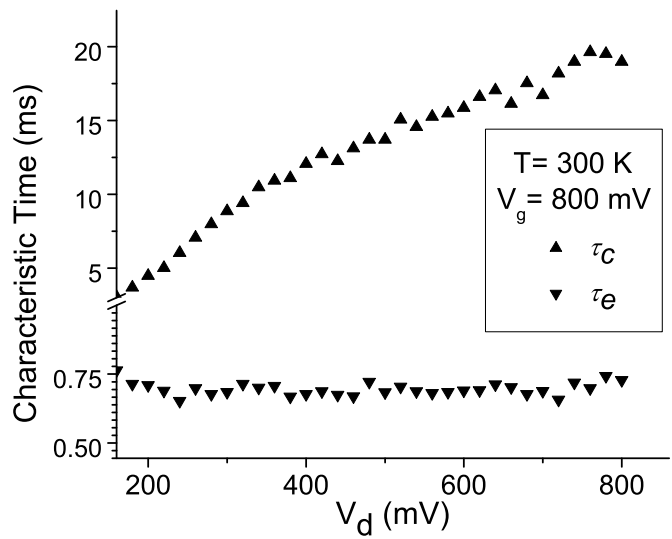
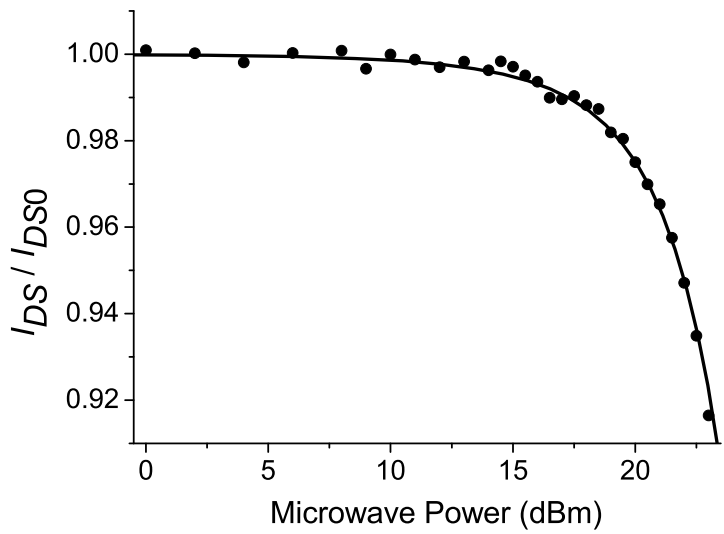
\end{document}